\documentclass[amsmath,amssymb,aps,pra,superscriptaddress,notitlepage]{revtex4-1}
\usepackage{graphicx,psfrag}
\usepackage[colorlinks=true,linkcolor=blue]{hyperref}
\usepackage{latexsym,mathrsfs,mathrsfs}
\usepackage{braket}
\usepackage{graphicx}
\usepackage{caption}
\usepackage{wrapfig} 
\usepackage[font=footnotesize,labelfont=bf,
justification=justified,
format=plain]{caption}
\usepackage{array}
\usepackage{lipsum} 
\usepackage{varioref}
\usepackage[utf8]{inputenc}
\usepackage{hyperref}
\hypersetup{colorlinks,%
citecolor=blue,%
filecolor=blue,%
linkcolor=blue,%
urlcolor=blue}
\usepackage{cleveref}

\begin{document}
\title{Experimental realisation of multipartite entanglement via quantum Fisher information in a uniform antiferromagnetic quantum spin chain}

\author{George Mathew}
\affiliation{School of Physics, IISER Thiruvananthapuram, Vithura, Kerala-695551, India}
\author{Saulo L. L. Silva}
\affiliation{Centro Federal de Educa\c c\~ao Tecnol\'ogica, CEFET-MG, Nepomuceno-37250000, Brazil}
\author{Anil Jain}
\affiliation{Solid State Physics Division, Bhabha Atomic Research Centre, Mumbai 400 085, India}
\affiliation{Homi Bhabha National Institute, Anushaktinagar, Mumbai 400094, India}
\author{Arya Mohan}
\affiliation{School of Physics, IISER Thiruvananthapuram, Vithura, Kerala-695551, India}
\author{D. T. Adroja}
\affiliation{ISIS Facility, Rutherford Appleton Laboratory, Chilton, Didcot, Oxon OX11 0QX, UK}
\affiliation{Highly Correlated Matter Research Group, Physics Department, University of Johannesburg, PO Box 524, Auckland Park 2006, South Africa}
\author{V. G. Sakai}
\affiliation{ISIS Facility, Rutherford Appleton Laboratory, Chilton, Didcot, Oxon OX11 0QX, UK}
\author{C. V. Tomy}
\affiliation{Department of Physics, IIT Bombay, Powai, Mumbai-400076, India}
\author{Alok Banerjee}
\affiliation{UGC-DAE CSR, Khandwa Road, Indore-452017, India}
\author{Rajendar Goreti}
\affiliation{School of Chemistry, IISER Thiruvananthapuram, Vithura, Kerala-695551, India}
\author{Aswathi V. N.}
\affiliation{School of Physics, IISER Thiruvananthapuram, Vithura, Kerala-695551, India}
\author{Ranjit Singh}
\affiliation{School of Physics, IISER Thiruvananthapuram, Vithura, Kerala-695551, India}
\author{D. Jaiswal-Nagar}
\email{deepshikha@iisertvm.ac.in}
\affiliation{School of Physics, IISER Thiruvananthapuram, Vithura, Kerala-695551, India}

\begin{abstract}
Quantum entanglement is a quantum mechanical phenomenon where the quantum state of a many body system with many degrees of freedom cannot be described independently of the state of each body with a given degree of freedom, no matter how far apart in space each body is. Entanglement is not only considered a resource in quantum information but also believed to affect complex condensed matter systems. Detecting and quantifying multi-particle entanglement in a many-body system is thus of fundamental significance for both quantum information science and condensed matter physics. Here, we detect and quantify multipartite entanglement in a spin 1$/$2 Heisenberg antiferromagnetic chain in a bulk solid. Multipartite entanglement was detected using quantum Fisher information which was obtained using dynamic susceptibility measured via inelastic neutron scattering. The scaling behaviour of quantum Fisher information was found to identify the spin 1$/$2 Heisenberg antiferromagnetic chain to belong to a class of strongly entangled quantum phase transitions with divergent multipartite entanglement. 
\end{abstract}

\maketitle

\section{Introduction}
Quantum information processing (QIP), namely, application of the laws of quantum mechanics for performing computations, is heralding the information revolution, wherein, intense theoretical and experimental investigations are underway to build quantum computers that can outsmart classical computers by using algorithms that can do computations much faster than a corresponding classical computer \cite{nielsen,boseprl}. In particular, entanglement \cite{connor,arnesan,gu,vedral_HightempEnt,amico,guhne,venutilong,wang,weisniak}, exemplified as a superposition of two spin $1/2$ states, ($1/\sqrt{2}$)($\ket{\uparrow\downarrow}$ - $\ket{\downarrow\uparrow}$), is used as a resource \cite{nielsen,boseprl} in the QIP task of quantum computation. \textbf{However, it is also known that if a given quantum system is not highly entangled, then it can be efficiently simulated on a classical computer itself \cite{vidal1,verstraete}. So, multipartite entanglement between qubits is of paramount importance for the functioning of a real quantum computer. In fact, the fragile quantum states produced during the task of quantum computation are protected from error by distributing a quantum state among many quantum bits using quantum entanglement \cite{barends,kelly}. Engineering a quantum system to perform computations in a large Hilbert space but with low error rates is the ultimate aim of real quantum computing platforms that are being developed by companies like Google, IBM, Righetti etc. \cite{gambetta,arute}. This requires cooling down the quantum processing system to very low temperatures (few mK) to protect the delicate quantum state from decoherence.}
Many measures to quantify bipartite entanglement, namely, entanglement between two subsystem of a given system, exist in the literature \cite{osborne,vidal,wooters,vedralprl,vedralrmp,gu,connor,arnesan,vedral_HightempEnt,amico,guhne,venutilong,roscilde,wang,weisniak}. Using such measures, bipartite entanglement has been reported in low dimensional spin systems, via specific heat and magnetization measurements \cite{brukner,sahling,souza,weisniak}. However, multipartite entanglement, where a pure state of a many-body system is not separable in any bipartite splitting \cite{hauke,hyllus,toth,gabbrielli,meyer,chandran,guhne,amico}, is extremely difficult to obtain in a many-body system. \textbf{Experimentally, many examples of multipartite entanglement generation in few-body systems have been reported, example, multipartite entanglement in six or eight trapped ions \cite{haffner} etc. The measurement of multipartite entanglement in such systems is done via protocols based on preparing many copies of the quantum system. So, measurement of multipartite entanglement in quantum many-body systems seems a virtually impossible task with an exponential scaling of resources.}\\
\textbf{A direct measurement of multipartite entanglement for many-body systems has been proposed by Hauke et al. \cite{hauke} in the form of quantum Fisher information (QFI), F$_Q$, a concept that originated in quantum metrology \cite{escher,toth}, where F$_Q$ defines the maximum precision with which a phase could be estimated given a quantum state. By relating QFI to Kubo response function, Hauke et al. proposed an experimental witness of multipartite entanglement in the form of QFI, F$_Q$ \cite{hyllus,toth,gabbrielli,hauke}- a witness of multipartite entanglement, via dynamic susceptibility \cite{hauke}. }To our knowledge, such a direct measurement of multipartite entanglement via QFI, has not yet been done on any many-body condensed matter system. \textbf{Multipartite entanglement is now believed to play an essential role in many complicated condensed matter phenomenon like high temperature superconductivity \cite{marel}, fractional quantum hall effect \cite{vedral_HightempEnt} and quantum phase transitions (QPT) that are phase transitions which happen at absolute zero of temperature when a non-thermal control parameter like magnetic field is tuned to a critical value called as a quantum critical point (QCP) \cite{vojta,osborne,gu,vidal}.}\\
Antiferromagnetic chains with global SU(2) invariance have rotational symmetry making them ideal candidates to test various features of multipartite entanglement \cite{vedral_HightempEnt,venutitele}. So, our model spin system that is used to detect multipartite entanglement is a spin $1/2$ antiferromagnetic Heisenberg chain (AfHc) which is the seminal and one of the simplest quantum many body systems \cite{bonner,bogoliubov,eggert,tennant,fledderjohann,starykhprb,starykhprl,klumper,hammar,johnston,lake,dpnas,mourigal,lakeprl,starykhphysicaB} whose exact ground state is a macroscopic singlet entangling all spins in the chain \cite{klumper,eggert,johnston,bogoliubov,fledderjohann,mourigal}. A spin 1$/$2 AfHc is also inherently quantum critical at \textit{T} = 0 K and zero applied magnetic field, and has a line of critical points with respect to an applied magnetic field \cite{bonner,bogoliubov,eggert,tennant,fledderjohann,starykhprb,starykhprl,klumper,hammar,johnston,lake,dpnas,mourigal,lakeprl}. Even though the QPT’s happen at absolute zero of temperature, the effect of the QCP is observed at low but finite temperatures. Vojta et al. \cite{vojta} argue that the limit of applicability of the \textit{T} = 0 QCP is governed by the microscopic energy scale of the problem. Using a spin $1/2$ uniform AfHc, we detect and quantify multipartite entanglement in the chain through quantum Fisher information obtained using the dynamical spin susceptibility measured via inelastic neutron scattering. To the best of our knowledge, this is the first experiment in any many-body spin system to unambiguously demonstrate multipartite entanglement. We find that the multi-partite entanglement not only exists at a very low temperature of 40 mK but also survives till a very high temperature of 6.7 K ($\sim$ 2.2 $J/k_B$), thus demonstrating the robustness of the spin $1/2$ antiferromagnetic chain as a very efficient many-body system that can sustain itself against decoherence. Simultaneously, it is also demonstrated that the many-body system belongs to the class of strongly entangled phase transitions with divergent entanglement and it is multipartite entanglement, rather than exchange energy, that governs the temperatures to which the effect of \textit{T} = 0 QCP is felt. 
\section{Experimental Details}
Single crystals of $[Cu(\mu-C_{2}O_{4})(4-$aminopyridine$)_{2}(H_{2}O)]_{n}$ were grown by a slow diffusion technique \cite{castillo}. Single crystal X-ray diffraction measurements were performed on a Bruker Kappa APEX II CCD diffractometer by the $\omega$ scan technique using graphite monochromatized Mo$K_{\alpha}$ radiation ($\lambda$ = 0.71073 $\AA$) at room temperature (293 K). Few single crystals of total mass 1.5 mg were aligned along the length of the needle for magnetization measurements. Magnetization measurements in the temperature range of 1.8 K to 300 K were done on a vibrating sample magnetometer attached to Quantum Design’s Physical Property Measurement System (Models 14 T PPMS-VSM and Evercool-II). Lower temperature measurements in the range of 0.49 K to 2 K were done on a Helium 3 insert attached to Quantum Design’s SQUID magnetometer (Model iHelium3). The inelastic neutron scattering data was collected at indirect geometry IRIS time-of-flight spectrometer using PG002 analyzer with elastic resolution of 17.5 $\mu$eV at ISIS, Rutherford Appleton Laboratory, U.K. Data at 40 mK and 500 mK temperatures were measured using a dilution refrigerator while that at 3 K and 6.7 K measured using a He-4 Orange cryostat. Since the mosaic spread of the large number of single crystals aligned for INS measurement was greater than 20$^{\circ}$, they were crushed to make powder form in order to enhance the signal to noise of the inelastic neutron scattering experiment. The obtained powder was put within an annular copper can with 20 mm diameter with He-exchange gas and connected to the above mentioned cryostats for the data acquisition. The data at each temperature were collected for $\sim$ a day. Prior to the measurement, the single crystals of $[Cu(\mu-C_{2}O_{4})(4-$aminopyridine$)_{2}(H_{2}O)]_{n}$ were deuterated by deuterating each of the reagents 4-aminopyridine and K$_2$[Cu(ox)$_2$].2H$_2$0 resulting in deuterated 4-aminopyridine (D4-Apy) and K$_2$[Cu(ox)$_2$].2D$_2$0 and then synthesising the crystals using the slow diffusion technique described as above. 
\section{Results}
\subsection{The model system and substance}
\begin{figure}
	\centering  
	\includegraphics[scale=0.55]{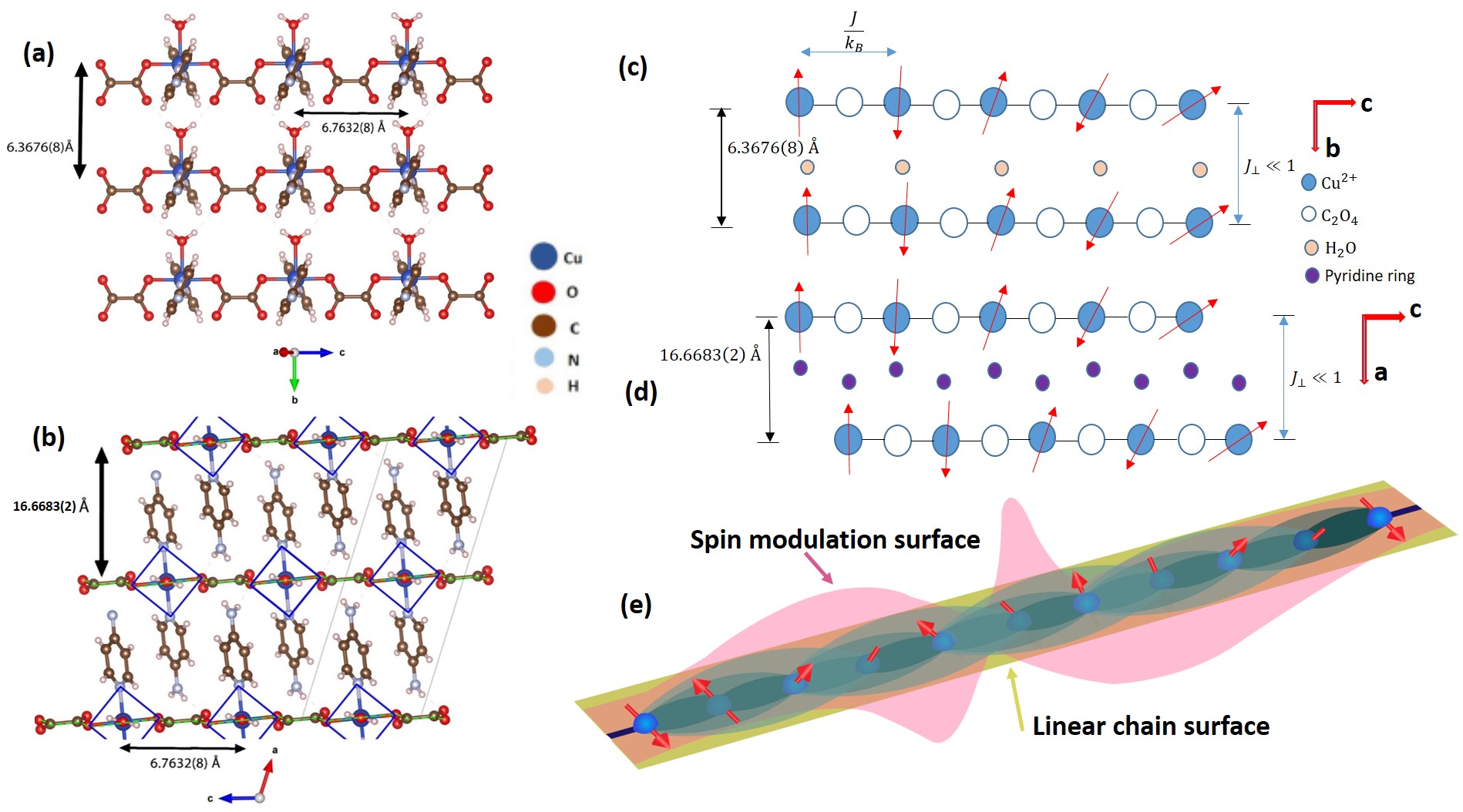} 
	\caption{(a): Crystal structure of $[Cu(\mu-C_{2}O_{4})(4-$aminopyridine$)_{2}(H_{2}O)]_{n}$. Chains of Cu$^{2+}$ ions (shown in blue) parallel to c-axis formed by bridging Cu$^{2+}$ ions with oxalate C$_2$O$_4$ molecules that co-ordinate to the Cu$^{2+}$ ions only through two of the oxygen atoms of the C$_2$O$_4$ unit, leaving the other two free. (b) Crystal structure in the a-c plane showing a distorted square-pyramidal co-ordination of Cu$^{2+}$ ion (shown by blue lines) formed by two oxygen atoms of the oxalate molecule and two nitrogen atoms of the pyridine ring. (c) and (d) show schematic representation of the crystal structure, wherein, spins carried out by the Cu$^{2+}$ ions are shown as red arrows. The magnetic interaction happens in the chain direction (c-axis) owing to the S = 1$/$2 spin carried by each Cu$^{2+}$ ion governed by the exchange coupling constant, $J/$k$_B$. Magnetic exchange in the other two directions, represented by J$_{\perp}$, is very small due to the presence of water molecule in the b-direction resulting in weak hydrogen bonds of the type O-H…O formed by the O-H of the $H_{2}O$ molecule and two free $O$ atoms of the oxalate molecule (see (a)), and the presence of pyridine rings in the a-direction resulting in a large separation of $\sim$ 16.6 $\AA$ between the Cu carrying linear chain. (e) Three dimensional spin representation on the linear Cu chain where the spins are shown to align antiferromagnetically and modulate on a surface according to Hamiltonian \ref{hamiltonian}. The entanglement between spins 1$/$2 is represented by green coloured contours with the nearest neighbor entanglement shown by the darkest shade of green. Next nearest and next to next neighbour entanglement is shown by progressively lighter shade of green. Such interactions exist between all atoms and are sketched for a few atoms for brevity.}
	\label{Fig:crystal structure}
\end{figure}

The spin 1$/$2 AfHc is modeled by the Heisenberg Hamiltonian describing nearest-neighbour interaction of localised quantum spins:
\begin{equation}
\label{hamiltonian}
H = J \sum_i \boldsymbol{S}_i \cdot \boldsymbol{S}_{i+1}, 
\end{equation}
where $J$ is the exchange coupling constant and $\boldsymbol{S}_i$ is the spin ope\-ra\-tor of the site $i$.\\
\textbf{The behaviour of bipartite entanglement in a spin $1/2$ AfHc has been investigated in quite a few studies \cite{connor,wooters,arnesan,gu,guhne,meyer,vedral_HightempEnt,venutitele,souza}. For instance, Connor et al. \cite{connor} calculated the pair-wise entanglement within a system of N particles arranged in a ring and found that each nearest neighbour pair tries to maximise the singlet state $1/\sqrt{2}$)($\ket{\uparrow\downarrow}$ - $\ket{\downarrow\uparrow}$) simultaneously. However, due to monogamy of entanglement, none of the pair reaches the maximum value of 1 \cite{connor}. On the other hand, Wie\'{s}niak et al. \cite{weisniak} demonstrated magnetic susceptibility to be a macroscopic witness of spin entanglement. For a spin 1$/$2 AfHc, the entanglement criterion was found to be:
\begin{equation}\label{EWineq}
\overline{\chi}^{expt}(T) = \frac{\chi^x(T)+\chi^y(T)+\chi^z(T)}{3} < \frac{(g\mu_B)^2 NS}{3k_B T},
\end{equation}
where $\overline{\chi}^{expt}(T)$ is the average of magnetic susceptibility measured along the three orthogonal directions, $g$ is the Land\'{e} $g$-factor, $\mu_B$ is the Bohr magneton and $N$ is the number of spin-$S$ particles. Using the equation \ref{EWineq}, a macroscopic witness of spin entanglement $MW_{SE}(T)$ is defined as \cite{souza}: 
\begin{equation}\label{EW}
MW_{SE}(T) = \frac{3k_B T\overline{\chi}^{expt}(T)}{(g\mu_B)^2 NS} - 1
\end{equation}
So, negative $EW(T)$ values detect macroscopic spin entanglement in the spin 1$/$2 AfHc chain}.\\  
The experimental realisation of such a spin $1/2$ AfHc is a blue coloured Cu$^{2+}$ containing co-ordination polymer $[Cu(\mu-C_{2}O_{4})(4-$aminopyridine$)_{2}(H_{2}O)]_{n}$ (abbreviated to CuP henceforth) \cite{castillo}, that crystallises from an aqueous solution as chains of Cu$^{2+}$ ions carrying spin $1/2$ in the c-direction interacting via an exchange coupling constant $J/$k$_B$ and a much weaker exchange in the $a$ and $b$ directions (see Figs. \ref{Fig:crystal structure} (a)-(d)).  \textbf{At $T$ = 0, the ground state consists of spins $1/2$ entangling into a macroscopic singlet \textbf{S}$_{total}$ = $\sum_n$S$_n$ = 0 \cite{klumper,eggert,johnston,bogoliubov,fledderjohann,mourigal}, such that each spin entangles with every other spin in such a way that multipartite entanglement as measured by global entanglement is maximum at 1 \cite{meyer}. The spin-spin correlation function decays algebraically, $\langle{S_0^{\alpha}}S_r^{\alpha}\rangle ~ \propto ~ \frac{(-1)^r}{r}$, $\alpha$ = x,y,z, indicating that the spin $1/2$ AfHc is inherently quantum critical due to long ranged quantum fluctuations  \cite{klumper,eggert,johnston,bogoliubov,fledderjohann,mourigal}, affecting the multipartite entanglement \cite{gu}. So, the spin $1/2$ AfHc is, in essence, similar to Anderson's resonating valence bonds (RVB), wherein, the spin singlet state is approximated by all possible RVB combinations each with a specified weight \cite{amico,venutilong}, the difference being that each given bond ``tries'' to be in the singlet state $1/\sqrt{2}$)($\ket{\uparrow\downarrow}$ - $\ket{\downarrow\uparrow}$) but doesn't achieve it due to monogamy of entanglement \cite{connor,wooters} unlike the RVB state where each bond is actually a singlet. RVB states have, in fact, been proposed to carry genuine multipartite entanglement over the entire lattice \cite{chandran}. So, a RVB like state in a spin 1$/$2 AfHc is expected to carry genuine multpartite entanglement. A snapshot of such multipartite entanglement is shown in Fig. \ref{Fig:crystal structure} (e) where the spin entanglement is schematically represented by green contours such that the strength of entanglement goes down as the distance from a given spin increases \cite{fledderjohann,bogoliubov}.}\\ 

\begin{figure}[hbtp]
	\centering  
	\includegraphics[scale=0.58]{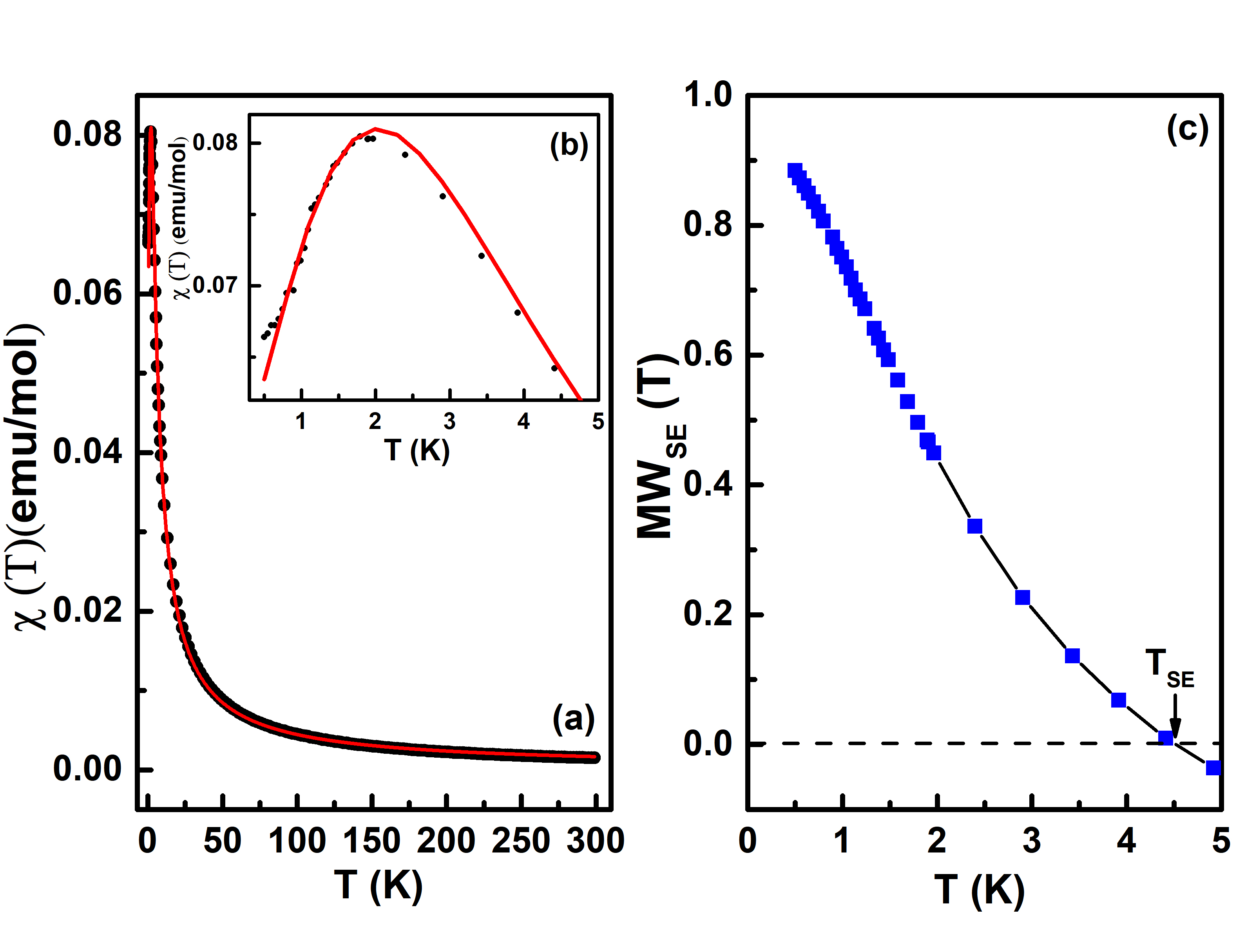} 
	\caption{(a) Black solid circles represent magnetic susceptibility, $\chi(T)= M (T)/H$, measured at an applied field $\mu_0$H = 1 T in the temperature range 0.49 K - 300 K. Red solid curve is a fit to the Bonner-Fisher model. (b) Expanded view of the data in (a) to zoom on the low temperature peak at $T^{max} = 1.95 K \pm 0.05 K$, obtained as the temperature where $d\chi/dT = 0$. (c) Filled blue squares correspond to the temperature variation of entanglement witness $MW_{SE}(T)$. Black solid line is a guide to the eye. $T_{SE}$ marks the temperature above which $MW_{SE}(T)$ becomes positive.}
	\label{susceptibility}
\end{figure}

In order to check that C$_{12}$H$_{14}$CuN$_4$O$_5$ (CuP) is an excellent representation of a spin 1$/$2 antiferromagnetic Heisenberg chain (AfHc), magnetization measurements were done on few aligned crystals as shown in Figs. \ref{susceptibility} (a) and (b) that show the clear presence of a low temperature peak characteristic of a spin 1$/$2 AfHc. The fitting to the experimental data was done using the expression $\chi(T)=C_{0}+C_{1}+\chi_{BF}(T)$, where $C_0$ is a small positive constant to account for a small $(0.4-1 \%)$ of uncoupled spin $1/2$ impurities and $C_{1}$ is the diamagnetic contribution from the closed atomic shells of CuP, estimated as $-16.7 \times 10^{-5} emu/mole$ by using tabulated values for Pascal’s constants \cite{abraham}. $\chi_{BF}(T)$ is the susceptibility given by the Bonner-Fisher model \cite{bonner}. The fit yielded a value of the exchange coupling constant $J/$k$_B$ as 3.1 K and a $g$ value of 2.1. From a rigorous high-temperature series expansion calculation \cite{johnston}, $T^{max}$ was obtained as 0.640851 $J/$k$_{B}$. Using the experimentally obtained value of $T^{max}$ = 1.95 K, a $J/$k$_{B}$ of 3.05 K was obtained, in excellent agreement to the one obtained from the Bonner-Fisher fit (3.1 K).   

In order to witness macroscopic spin entanglement on CuP using the entanglement witness $MW_{SE}(T)$ given by equation \ref{EW}, it is necessary to ensure that the minimal Hamiltonian (equation \ref{hamiltonian}) having isotropic Heisenberg exchange interactions (without any anisotropy) is able to capture the significant physics of the system. Our inelastic neutron scattering study shows no signature of long-range magnetic ordering down to 40 mK (see the next section). We also find no gap in the spin excitation spectrum ruling out the possibility of any significant anisotropy. This implies that the value of anisotropy, if any, is less than the resolution of the instrument ($\sim$ 17.5 $\mu$eV), which is much smaller than the value of nearest neighbour interaction $J$ $\sim$ 0.27 meV). We observed a multi-spinon continuum in the excitation spectra, which confirms that our system is an excellent representation of a spin 1$/$2 AfHc.\\
Fig. \ref{susceptibility} c shows macroscopic spin entanglement $MW_{SE}(T)$ obtained from the temperature dependence of magnetic susceptibility (Figs. \ref{susceptibility} (a) and (b)). It can be seen that $MW_{SE}(T)$ attains negative values till $T_{SE}$ impying that spin entanglement exists upto $T_{SE}$. However, positive $MW_{SE}(T)$ values above $T_{SE}$ do not necessarily imply separable states and bipartite entanglement between non-neighbouring sites or multipartite entanglement may occur between spins even at temperatures above $T_{SE}$ \cite{weisniak,souza}. So, “macroscopic witness of spin entanglement” using magnetic susceptibility has its limitations, in that, it may underestimate the temperature upto which spin entanglement may be present in a system and a genuine measure for multi-particle entanglement is multipartite entanglement.\\ 
\subsection{Inelastic neutron scattering data and quantum Fischer information}
\textbf{It is known that multipartite entanglement, where entanglement exists between more than two particles \cite{guhne,amico,hauke,gabbrielli,meyer,hyllus,chandran,toth}, offers many advantages in quantum tasks like noise metrology, quantum communication protocols etc. \cite{guhne,amico,meyer,toth,escher}. However, a direct detection of multipartite entanglement has not yet been done in any many-body condensed matter system. }So, in order to check if multipartite entanglement could contribute to positive $MW_{SE}(T)$ values above $T_{SE}$ in Fig. \ref{susceptibility} (c), we decided to try to detect multipartite entanglement between the spins in the AfHc. 
In this regard, Hauke et al. \cite{hauke} proposed quantum Fisher information \cite{hyllus,toth,gabbrielli}, F$_Q$(T), as a witness of  multi-particle entanglement in systems with large degrees of freedom, using dynamic susceptibility as follows:
\begin{equation}\label{QFI}
F_Q(T) = \frac{4}{\pi}\int_{0}^{\infty}d\omega~tanh\bigg(\frac{\omega}{2T}\bigg)~ \chi''(\omega,T)   
\end{equation}
 
where $\chi$''($\omega$, T) is the imaginary part of the dynamic susceptibility, $Q$ the wave-vector and $\omega$ the frequency. Since quantum fluctuations affect multipartite entanglement, it is natural that a correlation between $F_Q$(T) and dynamical susceptibility-a quantity that measures fluctuations, exists \cite{hauke}. In order to obtain $\chi$''($\omega$, T), we resorted to inelastic neutron scattering (INS)- a powerful technique that can provide direct information about the space and time Fourier transformation of the spin-spin correlation as function of energy and momentum transfer vector mapping over the entire Brillouin zone \cite{squires,lake,lakeprl,tennant} via the measured cross-section that yields the dynamical structure factor $S(Q,\omega)$. INS measurements were carried out at the time-of-flight (ToF) inverted-geometry crystal analyzer spectrometer IRIS at ISIS, UK, with fixed final energy of neutron 1.845~meV. The instrument was operated with an energy resolution (FWHM) of 17.5~$ \mu$eV, which allowed us to determine the possible presence of gap. 
\begin{figure}[hbtp]
	\centering
	\includegraphics[scale=0.8]{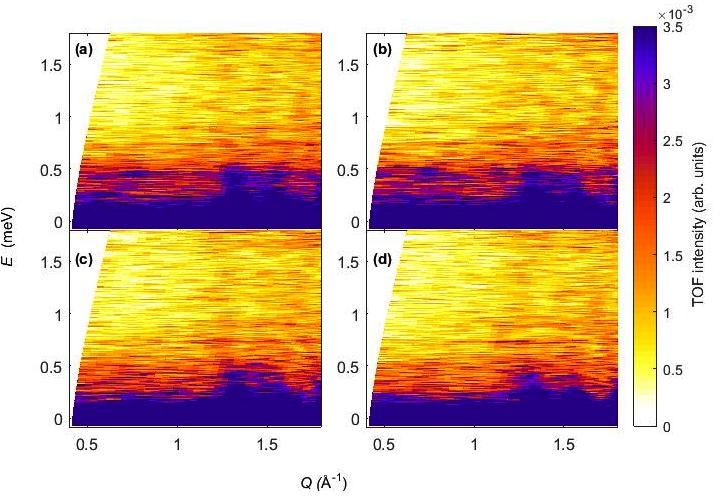}
	\caption{\label{fig:structure} (a)-(d) False colour representation of the inelastic neutron scattering (INS) spectra, S(Q,E), measured on a powder of crushed deuterated CuP single crystals of C$_{12}$D$_{14}$CuN$_4$O$_5$ at (a) 40 mK, (b) 500 mK, (c) 3~K, and (d) 6.7~K. TOF intensity is shown on the vertical scale bar.}
	\label{2Dplot}
\end{figure}
Figs. \ref{2Dplot} (a)-(d) show INS spectrum of deuterated CuP powdered sample measured at 40~mK, 500~mK, 3~K, and 6.7~K respectively. At 40 and 500~mK, the spectrum shows no sign of gap opening at the AFM zone center, $Q = 0.59~{\AA}^{-1}$, as expected for an ideal spin 1$/$2 Heisenberg spin chain system. A dispersive magnetic signal originating from AFM Zone center at $Q = 0.59~{\AA}^{-1}$, is clearly visible. The INS magnetic signal survives up to $\sim1$~ meV, however, its spectral weight decreases with increasing the energy transfer. Efforts to capture the observed feature in the INS spectra using the powdered average dispersion from linear spin wave theory by considering $J/k_B = 3.1$~K obtained from magnetic susceptibility measurements (see Fig. \ref{susceptibility} above) were not successful. 
 
\begin{figure}[h!]
	\centering  
	\includegraphics[scale=0.6]{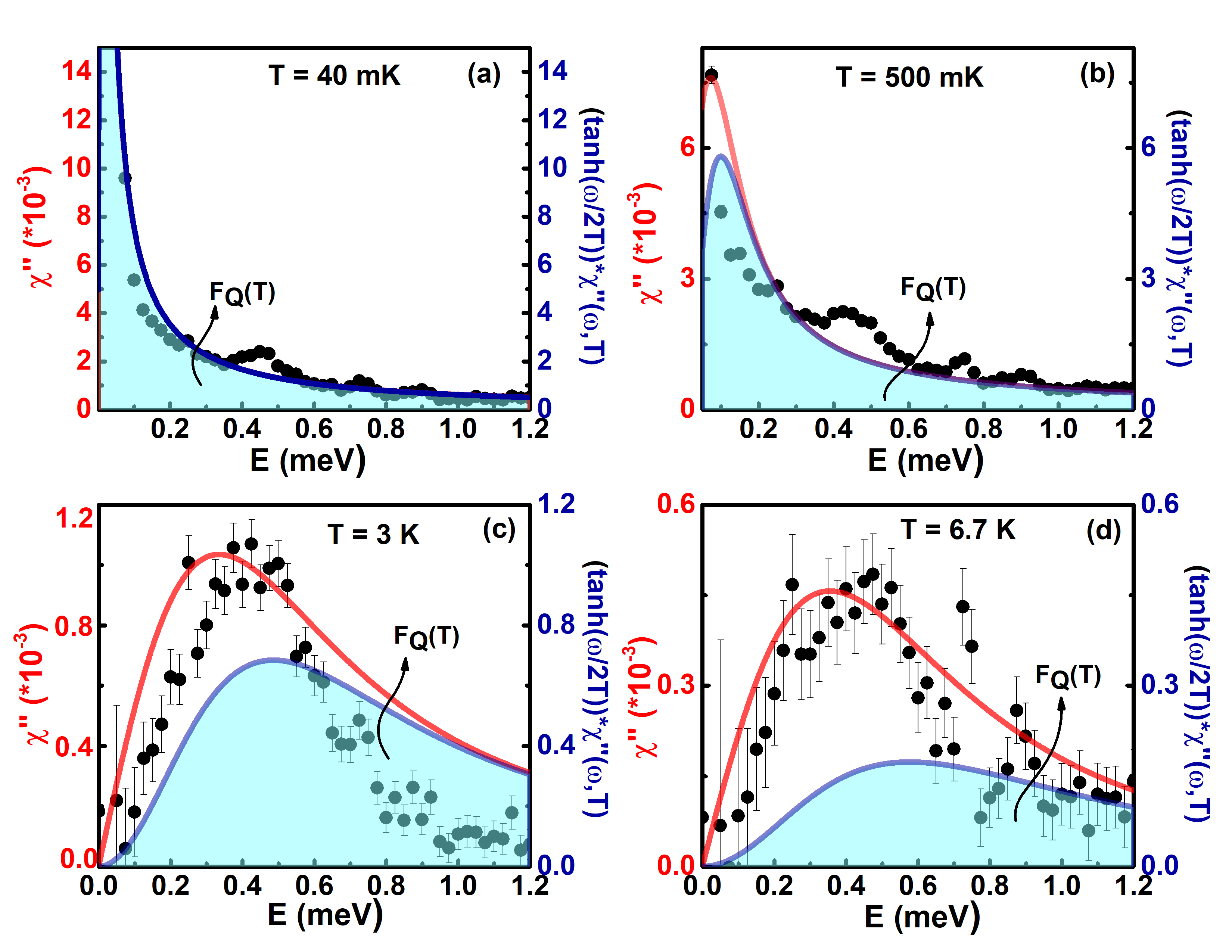} 
	\caption{Black filled circles in (a)-(d) represent the energy (\textit{E} = $\hbar \omega$) variation of imaginary part of dynamic susceptibility $\chi$''($\omega$) obtained from measured INS spectra at temperatures of 40 mK, 500 mK, 3 K and 6.7 K respectively after integration over momentum transfer Q from 0.4 to 1.1 $\AA ^{-1}$ and elastic line subtraction (see Fig. S4 of Supplementary Materials). Error bars in (a) and (b) are smaller than the size of the symbols. Red solid line in each panel is a fit to theoretical dynamical susceptibility calculated using equation \ref{starykh}, having $A_{Starykh}$ and $T_0$ as the fit parameters, and employing the fluctuation-dissipation theorem. The obtained best temperature independent fit parameters are $A_{Starykh}$ = 0.00065 and $T_0$ = $\pi$J/8. Blue solid curves in panels (a)-(d) represent (tanh($\omega/2T$))$\chi$''($\omega$,T). The cyan shaded area under the blue solid curve is $F_Q$(T) = (4/$\pi$)$\int_{0}^{\infty}$d$\omega$(tanh($\omega/2T$))$\chi$''($\omega$,T).}
	\label{Fisherinformation}
\end{figure}
The linear spin wave theory gives a magnon bandwidth of 0.55~meV, however, we observed magnetic signal up to 1~meV. The observed features in the excitation spectrum indicate the presence of spinon-like quasi particles, having two-spinon continuum, which could exist up to $\pi J$  which is the upper bound of two spinon continuum (see Appendix B), as expected for spin 1$/$2 one-dimensional Heisenberg AFMs. The spinon spectrum calculated for the INS data collected at 40 mK is described in the Appendix B. So, our inelastic neutron scattering study shows no signature of long-range magnetic ordering down to 40 mK as well as no gap in the spin excitation spectrum ruling out the possibility of any significant anisotropy. This implies that the value of anisotropy, if any, is less than the resolution of the instrument ($\sim$ 17.5 $\mu$eV), which is much smaller than the value of nearest neighbour interaction ($J$ $\sim$ 0.27 meV), confirming that our system is an excellent representation of a spin 1$/$2 AfHc.\\
\textbf{It was shown \cite{apollaro,wang} that the dynamics of a single spin chain can sustain the generation of two pairs of Bell states, so it is expected that the dynamics of many such spin chains in a spin 1$/$2 AfHc can together result in the generation of a large number of Bell states. Such a dynamics manifested in the low-lying excitations of a spin 1$/$2 AfHc, and known as spinons, define the dynamical structure factor.} The dynamics of a spin 1$/$2 AfHc recently calculated using conformal field theory and ``stochastic series expansion" quantum Monte-Carlo technique at low and intermediate temperatures in the form of dynamical structure factor \cite{starykhprl,starykhprb,starykhphysicaB} is given as:
\begin{widetext}
\begin{equation}\label{starykh}
S(Q,\omega) = \frac{1}{1- e^{-\hbar \omega/k_BT}}\frac{A_{Starykh}}{\pi T}2^{2\Delta -3/2} sin(2\pi \Delta) \left(ln\frac{T_0}{T}\right)^{1/2}\Gamma^2(1-2\Delta)Im\left(\frac{\Gamma^2(\Delta -i\frac{\omega}{4\pi T})}{\Gamma^2(1 - \Delta -i\frac{\omega}{4\pi T})} \right)
\end{equation}
\end{widetext}

where $\Delta = \frac{1}{4} \left(1-\frac{1}{2 ln \frac{T_0}{T}} \right)$ is a temperature dependent scaling dimension, $A_{Starykh}$ is a non-universal constant, $T_0$ is a high-energy cut-off, Im() is the imaginary part of the function inside the braces () and $\Gamma$ is the gamma function. The dynamical structure factor is converted, in the usual way, to imaginary part of dynamic susceptibility using the fluctuation-dissipation theorem \cite{squires,tennant} (see Appendix C). Since, the calculations for the dynamical structure factor in equation \ref{starykh} was done for low as well as intermediate temperatures, it is expected to work very well both at low temperatures (\textit{T} $<$ $J/k_B$) as well as high temperatures (\textit{T} $>$ $J/k_B$) \cite{starykhprl,starykhprb,starykhphysicaB} as is exemplified by the excellent fits of the theoretical curves (shown as red) to the experimentally obtained dynamical structure factor in Fig. \ref{Fisherinformation}. A small bump visible $\sim$ 0.45 meV in the 40 mK and 500 mK data may be arising due to the integration over the wave-vector Q from 0.4 to 1.1 $\AA ^{-1}$, involving the contribution from zone-bounday spinon as well as continuum from two spinon \cite{tennant,lakeprl} arising in a polycrystalline sample of ours. In contrast, in the cross-over region (T $\sim$ $J/k_B$), the fit of the theoretically obtained $\chi$'' to the experimentally obtained data is not that good, as expected. Quantum Fisher information, $F_Q$(T), is directly obtained from the area under the curve of (tanh($\omega/2T$))$\chi$''($\omega$,T), plotted as blue curves in Figs. \ref{Fisherinformation} (a) - (d). As far as we are aware, this is the first experimental proof of multipartite entanglement in any many-body system shown via quantum Fisher information. 
\section{Multipartite entanglement and quantum criticality}
\begin{figure}[hbtp]
	\centering  
	\includegraphics[scale=0.55]{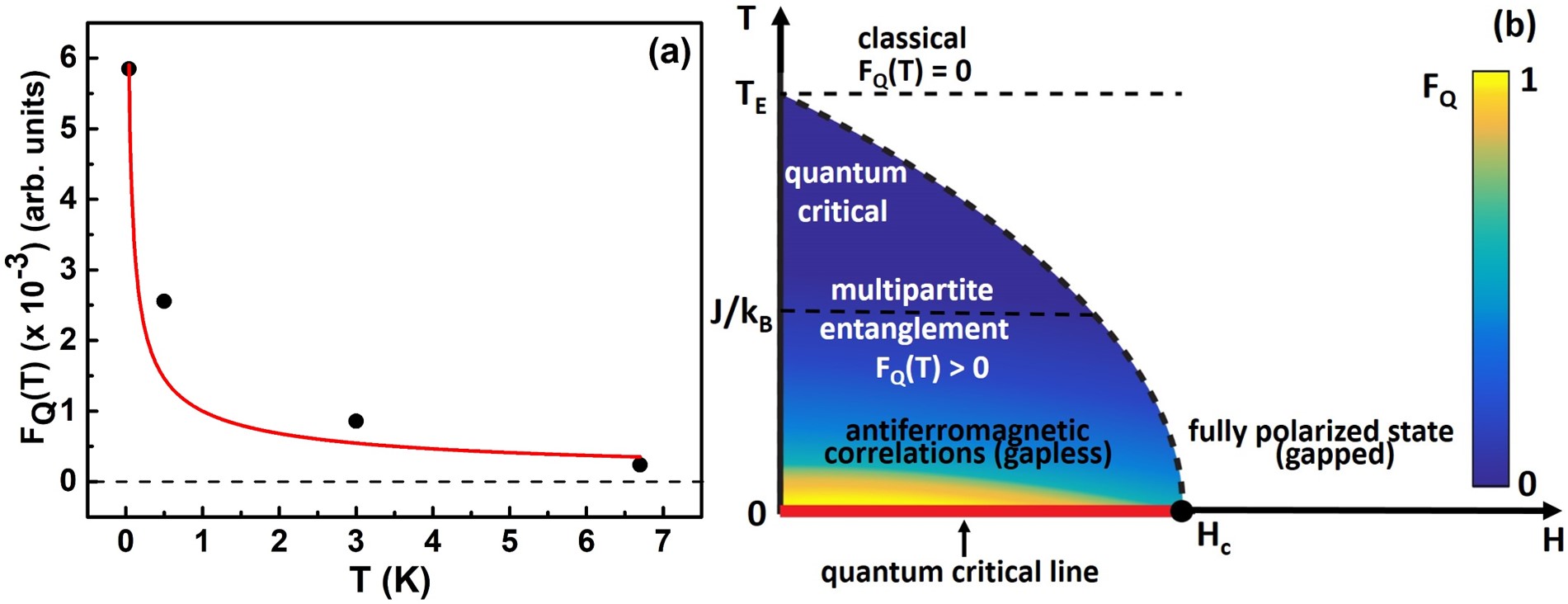} 
	\caption{(a) Black filled circles denote the temperature variation of quantum Fisher information, $F_Q$(T), calculated as described in the main text. Red solid curve is a fit according to equation \ref{scaling} where $\Delta_Q$, a fit parameter, was obtained as 0.55 and the dynamic critical exponent $z$ governing the magnetic field regime away from the critical end-point $H_c$ was taken as 1 \cite{klumper,eggert}. (b) Schematic phase diagram of a spin 1$/$2 antiferromagnetic Heisenberg chain that has a line of quantum critical points at $T$ = 0 (shown as a thick red line), where the horizontal axis denotes an applied magnetic field, $H$, and the vertical axis is the temperature \textit{T}. For a one dimensional spin 1$/$2 Heisenberg antiferromagnet, the quantum critical state at \textit{H} = 0 is that of an antiferromagnet that is continuously tuned with the application of a magnetic field and ends in a fully polarised state at a critical endpoint $H_c$ \cite{fledderjohann,bogoliubov,dpnas,hammar,lake}. Due to the existence of a line of critical points at \textit{T} = 0, a quantum critical region analogous to that of a cone for a single critical point, fans out to finite temperatures. Colour plots represent variation of quantum Fisher information, $F_Q$, inside the quantum critical region with blue representing minimum values and yellow denoting maximum values. Since the antiferromagnetic correlations decrease with an increase with magnetic field \cite{fledderjohann,bogoliubov,hammar}, a decrease of entanglement is expected for a spin 1$/$2 AfHc with magnetic field \cite{roscilde}$.~J/k_B$ has also been marked to show that $F_Q$ is small but still finite at $J/k_B$. $T_E$ marks the entanglement temperature at which temperature quantum Fisher information goes to zero and above which the system behaves classically.}
	\label{QC}
\end{figure}

Fig. \ref{QC} (a) presents the temperature variation of $F_Q$(T) obtained from equations \ref{QFI}, \ref{starykh} and fluctuation-dissipation theorem (see Appendix C).  At \textit{T} = 0.04 K, the AfHc is in a maximally entangled state. With an increase in the temperature, $F_Q$(T) starts to decrease due to the mixing of other states with the maximally entangled state decohering it \cite{arnesan,wang}. At the highest measured temperature of 6.7 K, multipartite entanglement, $F_Q$, is very small but still finite. It is expected to go to zero at an entanglement temperature $T_E$ above, but close to 7 K, which is significantly higher than that obtained in a spin dimer system Sr$_{14}$Cu$_{24}$O$_{41}$ where $T_E$ was found to be 2.1 K \cite{sahling}. This is an extraordinary result in view of the expectation that condensed matter systems having very large degrees of freedom would decohere easily due to interactions with the environment, inducing a quantum to classical transition \cite{nielsen}. Since the spin 1$/$2 AfHc has highly symmetric states having SU(2) global invariance \cite{venutilong}, it was expected that such states would exhibit a high degree of entanglement at high temperatures \cite{vedral_HightempEnt}. The existence of multipartite entanglement at very high temperatures of 6.7 K, then, confirms this expectation. Additionally, the temperature variation of $F_Q$(T) is seen to follow the scaling hypothesis \cite{hauke,gabbrielli} of equation \ref{scaling} very well, as shown by the red solid line in Fig. \ref{QC} (a): 
\begin{equation}\label{scaling}
F_Q \sim T^{-\Delta_Q/z}
\end{equation}
where $\Delta_Q$ is the scaling exponent of QFI and \textit{z} is the dynamic critical exponent. Hauke et al. \cite{hauke} proposed that systems exhibiting the scaling behaviour of equation \ref{scaling} can identify a class of strongly entangled QPT, namely, those QPT's that have a divergent QFI. These are those transitions that have $\Delta_Q >$ 0. Since $\Delta_Q$ was found to have a value of 0.55, a spin 1$/$2 AfHc belongs to the class of strongly entangled phase transitions with divergent multipartite entanglement.\\ 
In \cite{osborne,osterloh_QPT,gu}, the authors argued that QPT’s are genuinely quantum mechanical, in that, the property responsible for long range correlations is entanglement, such that the system is strongly entangled at the critical point. It is clear from the above discussions that quantum Fisher information can be used as a tool to identify strongly entangled quantum phase transitions. Such an identification has been done in the schematic representation of Fig. \ref{QC} (b) where multipartite entanglement is shown to govern the quantum critical region that extends to finite temperatures from \textit{T} = 0. At $T$ = 0 and $H$ = 0, $F_Q$ has a very large value that decreases as one moves to finite temperatures (c.f. Fig. \ref{QC} (a)). It is known that a spin 1$/$2 AfHc exhibits a line of continuously tuned quantum critical points in the \textit{H-T} plane by the application of a magnetic field that ends at the critical endpoint $H_c$ \cite{fledderjohann,bogoliubov,hammar,dpnas,lake}. So, the multipartite entanglement is expected to decrease with an increase in field, due to a decrease in the spin-spin correlations with field application \cite{fledderjohann,bogoliubov,hammar,roscilde}. Similar to the reduction of $F_Q$ at finite temperatures for the $H$ = 0 case, $F_Q$ is also expected to decrease at finite temperatures at finite fields as well. Finally, Vojta et al. \cite{vojta} proposed that the temperature at which universal behaviour associated with a critical point ceases to happen is governed by the characteristic energy scale of the problem, which is the exchange coupling constant in our case. It should be noted that in our spin 1$/2$ AfHc, multipartite entanglement governing quantum criticality exists at a temperature of 6.7 K which is more than double the exchange coupling constant ($J/$k$_B$ = 3.1 K) of our system. So, entanglement temperature T$_E$, rather than exchange coupling, marks the end of quantum criticality in a spin 1$/$2 AfHc.
\section{Summary and Outlook}
It has been shown that a spin 1$/$2 antiferromagnetic Heisenberg chain, the seminal model of a quantum many-body system, can detect multipartite entanglement via quantum Fisher information. Multipartite entanglement was detected using inelastic neutron scattering via dynamic susceptibility, the first experimental report of quantum Fisher information in any quantum many-body system. The multipartite entanglement was detected not only at the lowest measured temperature of 40 mK but was also found to survive to a very high temperature of 6.7 K, much higher than the exchange energy of this sytem. The obtained quantum Fisher information was also found to follow the scaling behaviour of systems corresponding to strongly entangled quantum phase transition. Since the AfHc has a high entanglement temperature, it is expected that the spin 1$/$2 antiferromagnetic Heisenberg chain will have a high entanglement teleportation fidelity at high temperatures \cite{nielsen,wang}, to enable our model spin 1$/$2 antiferromagnetic Heisenberg chain to be used as a bus to transmit quantum information at relatively high temperatures enhancing its technological scope manifolds.\\
\textbf{This study has revealed the synergy between quantum information science and condensed matter physics, wherein, the results of one area has been used to analyse and understand the details of the other and vice-versa. The complex ground state properties of a spin 1$/$2 uniform Heisenberg antiferromagnetic chain that contains all kinds of correlations were found to be of paramount importance in detecting and quantifying multipartite entanglement in it with possibilities of the antiferromagnet being used as a bus in a quantum computer which is a many-body system. It is hoped that similar studies will lead to deeper understanding of not only complicated condensed matter systems like high temperature superconductivity \cite{marel} but also areas of quantum information like quantum metrology \cite{escher,toth} etc.}   
      
\section*{Acknowledgements}
The authors thank G. Baskaran, Sreedhar Dutta, Naveen Surendran, S. Ramakrishnan, S. M. de Souza and M. Rojas for critically examining the manuscript and giving very valuable feedbacks. The authors thank Winfried Kockelmann for his help in performing neutron diffraction experiments on GEM instrument to check if the crystals were deuterated properly and David Voneshen for his help in checking the quality of the crystals on the ALF instrument. GM and DJ-N acknowledge Newton funding travel support for visiting ISIS Rutherford Appleton Laborating for the inelastic neutron scattering experiments.

\section*{Appendix: Deuteration of single crystals, spinon spectrum extraction and fluctuation-dissipation theorem}
\subsection{Deuteration of the single crystals}
\begin{figure}[hbtp]
	\centering
	\includegraphics[scale=0.51]{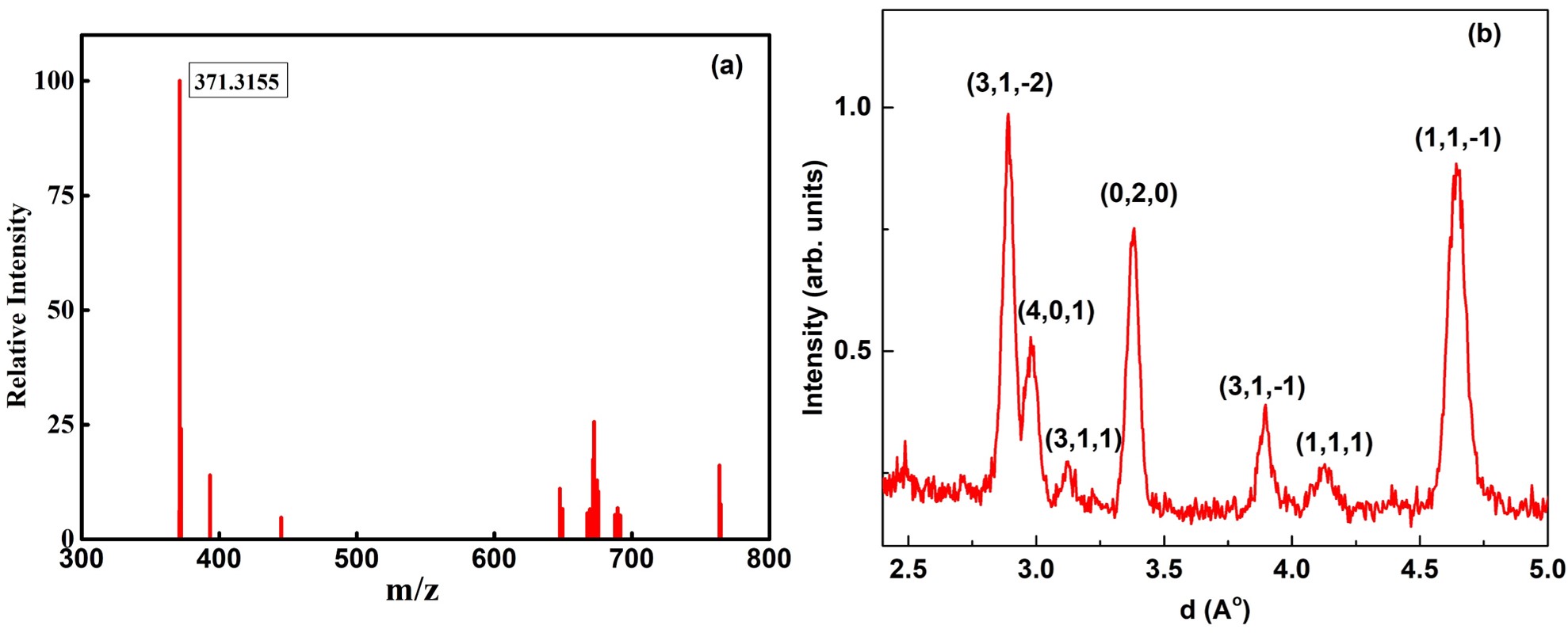}
	\caption{(a) High resolution mass spectra obtained on a single crystal of C$_{12}$D$_{14}$CuN$_4$O$_5$, taken in the range of 150 to 2000 m/z using Thermo Fischer scientific's Q Exactive TM- Bench top- LC-HRMS spectrometer. (b) Red curves correspond to neutron diffraction data measured on many co-aligned crystals totaling a mass of 0.5 g on the GEM time-of-flight neutron diffractometer at ISIS, Rutherford Appleton Laboratory, U.K. The shown spectra is a representative spectra measured on bank 3. Similar spectra were obtained on banks 1 and 2.}
	\label{HRMS and diffraction}
\end{figure}

It is a very well known fact that hydrogen has a very high incoherent  scattering cross-section with neutrons \cite{ramirez,shu} leading to high background and bad signal-to-noise in the inelastic neutron scattering experiments. In order to avoid this, crystals of $[Cu(\mu-C_{2}O_{4})(4-$aminopyridine$)_{2}(H_{2}O)]_{n}$, molecular formula C$_{12}$H$_{14}$CuN$_4$O$_5$ and molecular weight 357.8118 mol/g, were deuterated to $[Cu(\mu-C_{2}O_{4})(D4-$aminopyridine$)_{2}(D_{2}O)]_{n}$, molecular formula C$_{12}$D$_{14}$CuN$_4$O$_5$ and molecular weight 371.9238 mol/g, by first deuterating the starting reagents, namely, aminopyridine to deuterated 4-aminopyridine and K$_2$[Cu(ox)$_2$].2H$_2$O to K$_2$[Cu(ox)$_2$].2D$_2$O. Details will be published elsewhere \cite{george}. In order to confirm that the deuteration took place, we did a liquid chromatoraphy-high resolution mass spectroscopy (LC-HRMS) on a single crystal of C$_{12}$D$_{14}$CuN$_4$O$_5$ as shown in Fig. \ref{HRMS and diffraction} (a). A main peak at 371.3155 corresponding to a fragment with m/z 371.3155 can be clearly seen in the LC-HRMS spectra, indicating that H/D exchange at the H site of C$_{12}$H$_{14}$CuN$_4$O$_5$ has taken place.

In order to further confirm that the crystals were deuterated, they were subjected to neutron diffraction. The obtained neutron diffraction spectra is shown in Fig. \ref{HRMS and diffraction} (b) where the peaks were indexed according to the crystal structure obtained for C$_{12}$H$_{14}$CuN$_4$O$_5$ since single crystal x-ray diffraction cannot differentiate between H and D atoms. However, the flat background obtained in the neutron diffraction data of Fig. \ref{HRMS and diffraction} (b) is a clear indication that H/D exchange has taken place since H atoms scatter neutrons incoherently resulting in background scattering and, consequently, an extremely noisy background \cite{ramirez,shu}.

\subsection{Spinon spectrum}

\begin{figure}[hbtp]
	\centering
	\includegraphics[scale=0.6]{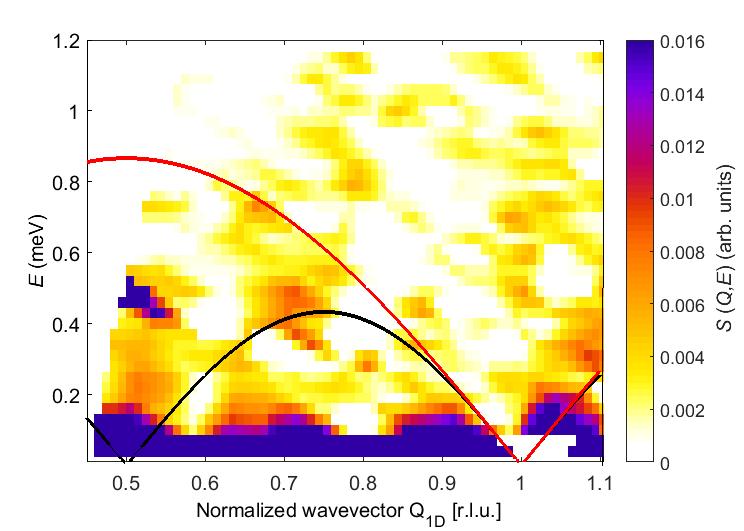}
	\caption{Single crystal like data, S$_{sx}^{(1D)}$(Q$_{1D}$,E), showing the spinon spectrum of our spin 1$/$2 AfHc, obtained by applying the conversion method using equations \ref{spinon extraction1} and \ref{spinon extraction2} to the INS data at 40 mK. Red and black solid lines represent the lower and upper bounds of the spinon continuum. Colours indicate the neutron scattering cross-section, S(\textbf{Q},E)}.
	\label{spinon}
\end{figure}
In order to get information about the spin dispersion of our system, the polycrystalline INS data presented in Fig. \ref{2Dplot} (a) were converted to a single-crystal-like data as shown in Fig.~\ref{spinon}, using the conversion method suggested by Tomiyasu et al.~\cite{tomiyasu,leiner}, applicable for one dimensional systems. According to this method, the powder average scattering function S$_{pwd}$(Q,E) (where Q = $|$\textbf{Q}$|$,E) is related to the single crystal scattering function S$_{sx}$(\textbf{Q},E) as:
\begin{equation}\label{spinon extraction1}
S_{pwd}(Q,E) = \int_{\textbf{Q} = Q}d\Omega S_{sx}(\textbf{Q},E)/(4\pi)
\end{equation}
where $\Omega$ is the solid angle over which the integral is done. For one dimension, S$_{sx}$(\textbf{Q},E) reduces to S$_{sx}^{(1D)}$(Q$_{1D}$,E), where
\begin{equation}\label{spinon extraction2}
S_{sx}^{(1D)}(Q_{1D},E) = [\partial QS_{pwd}(Q,E)/\partial Q]_{Q = Q_{1D}}
\end{equation}

Using the extraction method, the obtained  excitation spectrum Q-E for deuterated CuP is displayed in Fig.~\ref{spinon}. It is to be noted that in an inelastic neutron scattering experiment, only a pair of spinon can be observed because of the selection rule associated with a change in neutron spin by 1 in magnetic INS. Our experimental data clearly show a continuum between the lower and upper spinon two-spinon boundaries given by equations \ref{lowerspinon} and \ref{upperspinon} respectively:
\begin{equation}\label{lowerspinon}
E_l(Q_{1D}) = \frac{\pi J}{2}\big\vert sin(Q_{1D}c)\big\vert
\end{equation}

\begin{equation}\label{upperspinon}
E_u(Q_{1D}) = \pi J\bigg\vert sin\bigg(\frac{Q_{1D}c}{2}\bigg)\bigg\vert
\end{equation}

where $c$ is the lattice parameter along the chain direction. The value of the exchange coupling constant, $J/$k$_B$, was obtained from magnetic susceptibility measurements (see Fig. \ref{susceptibility}). So, it is very satisfying to observe that the spinon boundary obtained using the exchange coupling constant from magnetization measurements fit our INS data very well.

\begin{figure}[hbtp]
	\centering
	\includegraphics[scale=0.6]{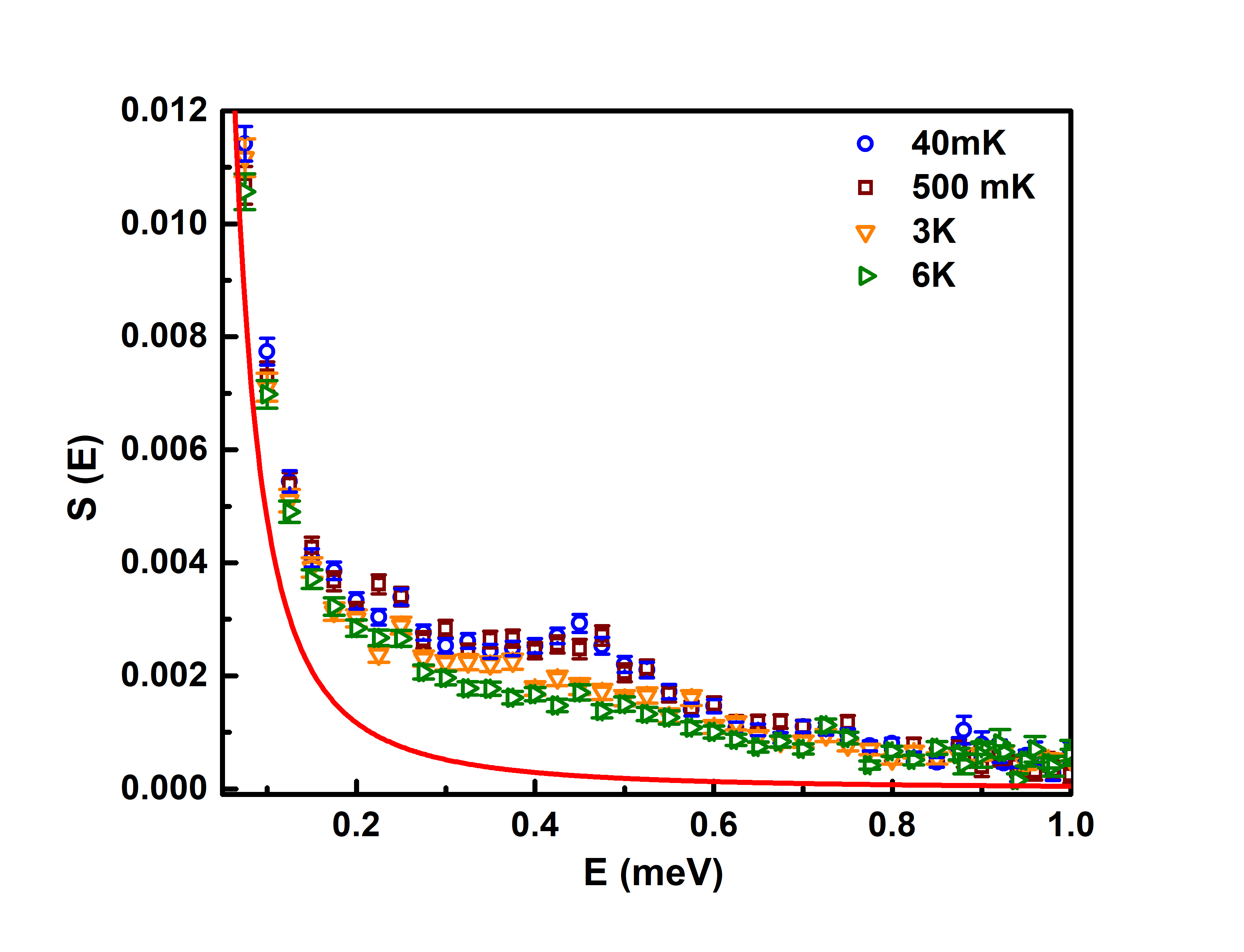}
	\caption{Raw inelastic neutron scattering data measured at ISIS, Rutherford Appleton Laboratory, U.K. Blue, wine, orange and olive coloured symbols denote the data points measured at temperatures of 40 mK, 500 mK, 3 K and 6.7 K respectively. Red solid curve is the resolution and background contribution obtained using the FWHM of standard vanadium sample.}
	\label{background}
\end{figure}

To capture the essential feature of the spinon spectrum, especially intensity modulation and gap, the INS data (40~mK, 500~mK, 3~K, and 6.7~K) have been integrated over momentum transfer \emph{Q} from 0.4 to 1.1 $\AA ^{-1}$ and plotted in Fig.~\ref{background} as a function of energy. The elastic line background, shown as solid red line in Fig.~\ref{background}, was subtracted to obtain $\chi$''($\omega$, T).

\subsection{Fluctuation-dissipation theorem}
The dynamical structure factor, $S(Q,\omega)$, for a spin 1$/$2 AfHc was obtained as \cite{starykhprl,starykhprb,starykhphysicaB}:
\begin{equation}\label{starykh}
S(Q,\omega) = \frac{1}{1- e^{-\hbar \omega/k_BT}}\frac{A_{Starykh}}{\pi T}2^{2\Delta -3/2} sin(2\pi \Delta) \left(ln\frac{T_0}{T}\right)^{1/2}\Gamma^2(1-2\Delta)Im\left(\frac{\Gamma^2(\Delta -i\frac{\omega}{4\pi T})}{\Gamma^2(1 - \Delta -i\frac{\omega}{4\pi T})} \right)
\end{equation}

In order to convert $S(Q,\omega)$ to imaginary part of dynamic susceptibility, $\chi''(Q,\omega)$, we used the fluctuation dissipation theorem \cite{squires,tennant}: 
\begin{equation}\label{chi''}
S(Q,\omega) = \left(1-e^{-\frac{\hbar \omega}{k_BT}}\right)^{-1}\chi''(Q,\omega)
\end{equation}


\bibliographystyle{unsrt}
\setcitestyle{square}

\end{document}